\documentclass[aps,twocolumn]{revtex4}
\usepackage{epsfig}
\usepackage{color}
\usepackage[retainorgcmds]{IEEEtrantools}

\begin{document}

\title{Effect of diffusion in  one-dimensional discontinuous 
absorbing phase transitions}

\author{Carlos E. Fiore} 

\email{fiore@if.usp.br} 

\author{Gabriel T. Landi} 

\email{gtlandi@gmail.com}

\affiliation{Instituto de F\'{\i}sica,
Universidade de S\~{a}o Paulo, \\
Caixa Postal 66318\\
05315-970 S\~{a}o Paulo, S\~{a}o Paulo, Brazil}

\date{\today}

\begin{abstract}

It is known that   diffusion provokes substantial changes in
 continuous absorbing phase transitions. 
Conversely,  its effect on discontinuous transitions is much less
 understood. 
In order to shed  light in this direction, we 
study the inclusion of diffusion in the simplest 
one-dimensional model with a discontinuous absorbing phase 
transition, namely the long-range contact process ($\sigma$-CP). 
Particles interact as in the usual CP, but the transition 
rate depends on the length $\ell$ of inactive sites 
according to $1 + a \ell^{-\sigma}$, where $a$ and $\sigma$ are control parameters. 
%In the absence of diffusion, this system presents both a discontinuous and a continuous phase transition, depending on the value of $\sigma$.
The inclusion of diffusion in this model has been investigated 
by numerical simulations and  mean-field calculations. 
Results show that there exists three distinct regimes. For 
sufficiently low and large $\sigma$'s the transition is respectively  always 
discontinuous or continuous, 
 independently of the strength of the diffusion. 
On the other hand, in an intermediate range of $\sigma$'s, the 
diffusion causes a suppression of the  phase coexistence
leading to a continuous transition belonging to the DP universality class.
%This  set of results does not agree with mean-field predictions, whose reasons will be discussed further.
\\
PACS numbers: 05.70.Ln, 05.50.+q, 05.65.+b

%It is argued that the inclusion of diffusion may provoke substantial changes 
%in second-order absorbing phase transitions. On the other hand, the situation
%is much less understood for discontinuous transition ones. 
 %In order to shed some light in this direction,  we analyze the 
%effect of diffusion 
%in the simplest example of an one-dimensional discontinuous
%absorbing transition. In the model, 
%(a long-range version of the CP) 
%particles interact as in the usual CP, but the transition
%rate depends on  the length $\ell$
%of   inactive islands surrounding particles according
%to relation $1+a\ell^{-\sigma}$, where and $a$ and $\sigma$ are
%control parameters. 
%The effect of diffusion has been investigated thoroughly for  
%remarkable values of $\sigma$.
%Our results show that  the diffusion  shortens the
%coexistence  line, which moves toward lower values of $\sigma$, in which
%appearing three distinct scenario
%depending on the value of $\sigma$.  
%For sufficient small $\sigma$'s,
%the transition remains first-order,
%even  for sufficient strong diffusion rates. On the other hand, 
%for intermediate values
%of $\sigma$'s,  phase coexistence is suppressed by the
%increase of diffusion, which becomes
%second-order belonging to the directed percolation (DP)
%universality class.  Finally,
%for larger $\sigma$'s, where the long-range interaction
%does not play important role and originally the phase transition
%is continuous, the diffusion does not provoke
%qualitative changes and the transition remains
%continuous for all diffusion rates.
 %
\end{abstract}

\maketitle

%-------------------------------------------------------
\section{Introduction}

Discontinuous absorbing phase transitions in low dimensions
have attracted a great deal of interest in recent years
\cite{marr99,ziff86,fiore04,hinr00,gine05,fiore2013,fiore2014}. 
Much of this effort has been based on the fundamental problem of
determining the  necessary 
ingredients for their occurrence.
Generically, discontinuous transitions  require an effective mechanism that 
suppresses the formation of absorbing minority islands (within
the active phase) induced by fluctuations. 
Although they may occur  in larger dimensions, 
there are strong evidences  that in one dimension short
range interactions cannot  stabilize compact clusters  \cite{hinr00}. 
In contrast,
a long-range counterpart of the contact process (CP), named
$\sigma-$CP \cite{gine05}, has revealed 
such occurrence, even in one dimension.
In the $\sigma-$CP,   particles are  created and annihilated like
in the usual short-range CP \cite{marr99}, but
the creation rate depends on the length $\ell$ 
of the island of inactive sites  according to the expression
$1+a\ell^{-\sigma}$.  For small $\sigma$ the interactions are  effectively 
long-range leading to a discontinuous transition \cite{gine05}. 
This can be understood by noting that the  long-range interaction
 introduces a collective behavior that is able to 
suppress the formation of minority islands. On the other hand,
when $\sigma$ is large 
the phase transition becomes  continuous and belongs to the 
directed percolation (DP) universality class, in similarity
with the usual CP.
Hence  there exists a tricritical point  
$\sigma_t$ where the 
transition changes from discontinuous to continuous. 

It is known that the presence of certain dynamics, such 
as disorder and diffusion, may drastically change the critical behavior 
and the classification of the phase transition.   
As in the equilibrium case, disorder may induce distinct 
universality classes and
Griffiths phases \cite{marr99,odor04,henkel,vojta}. 
Particle diffusion, on the other hand, can  also  be 
responsible for substantial changes, 
including not only the emergence  of distinct universality 
classes \cite{henkel} but also  
the  appearance of novel structures in the phase diagrams  \cite{dic89}.
However, much of the effort on understanding the role of these
ingredients  has focused    on continuous transitions, with the 
situation for discontinuous transitions
being much less understood \cite{munoz2014}. 

 In order to shed some light in this direction,  we analyze the 
role of diffusion in the $\sigma$-CP model, which is perhaps  
the  simplest  one-dimensional system
presenting a discontinuous absorbing transition.
This has been accomplished using numerical simulations performed 
in both the constant rate (ordinary)  \cite{marr99}
and the constant particle number (conserved) 
ensembles \cite{tome01,hilh02,oliv03,fiore07}. 
The simulations were performed for several values 
of the diffusion rate and several values of $\sigma$. 
We have found that the tricritical point $\sigma_t$ (which signals the crossover from continuous to discontinuous transition)
decreases as the diffusion increases. 
This leads to three distinct scenarios. For sufficiently 
small and large values of $\sigma$, the diffusion does not 
change the phase     
transition, remaining always  discontinuous and continuous, respectively. 
On the other hand, in an intermediate range of $\sigma$'s, 
the diffusion causes a suppression of the  phase coexistence
leading to a continuous transition belonging to the DP universality class.
The problem is also studied using mean-field theory in the pair approximation. 
As we show, however, mean-field calculations do not agree with the numerical results: they predict
that $\sigma_t$ should increase with $D$ and not the inverse. 
The reasons for this discrepancy are discussed. 

This paper is organized as follows: In Sec. II we present the model and 
in Sec. III the mean-field results. Numerical results are presented
in Sec. IV and in Sec. V we draw the conclusions.

%--------------------------------------------------------
\section{Model}

The one dimensional diffusive $\sigma-$CP 
\cite{gine05}  is defined as follows. To each site $i$ of a one-dimensional
lattice we associate an occupation variable $\eta_i$ that takes
the values 0 or 1 according to  whether the site is empty or
occupied. The dynamics involves three processes: spontaneous annihilation of a single particle (schematically
represented by $1\to 0$),
catalytic creation of a particle ($0\to 1$) and particle
hopings to a nearest-neighbor empty site 
($01\to 10$ or $10\to 01$).  The transition rate $w_i$
is given by the following expression:
\begin{equation}\label{rates}
w_i = D w_{i,i+1}^D+(1-D)(\alpha w_i^a+w_i^c),
%w_i = D\omega_{i,i+1}^D+(1-D)(\alpha \omega_i^a+\omega_i^c).
\label{eqor}
\end{equation}
where $D$ and $\alpha$ are the diffusion 
and annihilation rates, respectively and  $w_{i,i+1}^D$, $w_i^a$ and $w_i^c$ represent the diffusion,
annihilation and creation processes respectively. 
Their expressions are given by $w_{i,i+1}^D=\eta_i {\bar\eta}_{i+1}+{\bar\eta_i} \eta_{i+1}$
(where $\bar{\eta}_i = 1 - \eta_i$), $w_i^a= \eta_i$ and
\begin{IEEEeqnarray}{rCl}
\label{wic}w_i^c 
&=\frac12 
\sum_{\ell=1}^\infty  (1+a \ell^{-\sigma} )&\Big\{\eta_{i-1} {\bar\eta}_i
{\bar\eta}_{i+1}\ldots{\bar\eta}_{i+\ell-1}\eta_{i+\ell}  \\[0.2cm]
&&+ \eta_{i+1}{\bar\eta}_i  {\bar\eta}_{i-1}\ldots{\bar\eta}_{i-\ell+1}\eta_{i-\ell} \Big\},\nonumber
\end{IEEEeqnarray}
where $a$ and $\sigma$ are
parameters. When $a=0$ one recovers the original short-range
CP \cite{marr99,harr74}. 
As in that case, single particles
are created  only in  empty sites surrounded by at least one particle.
However, in the $\sigma-$CP the  creation rate
depends on the length $\ell$ between the  particles surrounding
the empty chosen site. 

For large values of $\alpha$, the system is constrained into the
absorbing state, in which no particles are allowed to be created.
Decreasing the parameter $\alpha$, a phase transition to 
an active state takes place, whose  location  and classification depends
on the parameters $a$, $\sigma$ and $D$. In order
to compare with previous results \cite{gine05,fiore07,fiore2013}, we 
 take the  value $a=2$. In this case, when $D = 0$ 
the crossover regime occurs at $\sigma_t = 1.0(1)$; i.e., 
the transition is continuous for $\sigma>1$ (belonging 
to the DP
universality class) and discontinuous for  $0<\sigma<1$.

%-------------------------------------------------------
\section{Mean field results}
Here we present a mean-field analysis of the $\sigma$-CP with diffusion.
Let $f(\eta)$ be an arbitrary function of the vector $\eta = (\eta_1,\eta_2,\ldots)$ 
 and define $\eta^i = (\eta_1,\eta_2,\ldots,\bar{\eta}_i,\ldots)$ and $\eta^{i,j} = (\eta_1,\eta_2,\ldots,\bar{\eta}_i,\ldots,\bar{\eta}_j,\ldots)$. From the underlying master equation it can be shown that the equation governing the time 
evolution of $\langle f \rangle$ is 
\begin{eqnarray}
%{rCl}
\frac{df}{dt} &=& \sum\limits_{i=1}^N \left\{ D \left\langle [f(\eta^{i,i+1})-f(\eta)] \omega_{i,i+1}^D \right\rangle \right. 
\\[0.2cm]
&&\left. + (1-D) \left\langle [f(\eta^{i})-f(\eta)] [\omega_{i}^c+\alpha\omega_i^a] \right\rangle \right\},
\label{expectation}
\end{eqnarray}
where the first and second terms take into account the particle diffusion and
the  creation and annihilation subprocesses, respectively.

From Eq.~(\ref{expectation})  we derive relations from 
the mean-field approach in the pair approximation. 
This second order approximation is required since, in the simple mean-field, the diffusion terms drops out entirely. 
Hence, we approximate the probability pertaining to a 
string of sites by 
\begin{equation}\label{pair_approx}
P(\eta_1,\eta_2,\eta_3,\ldots,\eta_\ell) \simeq \frac{P(\eta_1,\eta_2) P(\eta_2,\eta_3) \ldots P(\eta_{\ell-1},\eta_\ell)}{P(\eta_2)P(\eta_3)\ldots P(\eta_{\ell-1})}.
\end{equation}
Since two equations are required, we chose the system density  $\rho = P(1) = \langle \eta_i \rangle$ and 
 the two-site probability 
$z = P(01) = \langle \bar{\eta}_i \eta_{i+1} \rangle$.
Making use of the translation symmetry of the problem it is possible to show that Eq.~(\ref{expectation}) for $f = \eta_i$ becomes
\begin{equation}\label{MF_1}
\frac{d\rho}{d t} = (1-D) \left\{ z - \alpha \rho + \frac{a z^2}{1-\rho-z} \text{Li}_\sigma \left(\frac{1-\rho-z}{1-\rho}\right) \right\},
\end{equation}
where $\text{Li}_\sigma$ is the PolyLog function defined as 
\[
\text{Li}_\sigma(x) = \sum\limits_{\ell = 1}^\infty \frac{x^\ell}{\ell^\sigma}.
\]
Similarly Eq.~(\ref{expectation}) for $f = \bar{\eta}_i \eta_{i+1}$ yields
\begin{eqnarray}
%{rCl}
\label{MF_2}
\frac{dz}{dt} &=& (1-D)\left\{ \alpha(\rho-2z) - \frac{(1+a)z^2}{1-\rho}\right\}  \\[0.2cm]
&&+ 2D \left\{z - \frac{z^2}{\rho(1-\rho)}\right\}.\nonumber
\end{eqnarray}
 The steady-state behavior is obtained by setting $d\rho/dt = dz/dt=0$ in 
Eqs.~(\ref{MF_1}) and (\ref{MF_2}), which yields a system of algebraic equations for $\rho$ and $z$
\footnote{ It 
is possible to solve Eq.~(\ref{MF_2}) for $z$ as a function 
of $\rho$ and then insert this result into Eq.~(\ref{MF_1}). 
This has the advantage of yielding a single algebraic equation 
for $\rho$ as a function of the parameters $a$, $D$, 
$\sigma$ and $\alpha$, which is much simpler 
to solve numerically than a system of two coupled equations.}. 

In resemblance with the ``wan-der-Waals''   loops, that signals a discontinuous transition studied under mean- field like approaches, the phase transitions can be identified by the existence of a spinodal behavior. The co-existence point is estimated by the maximum value of $\alpha$, whose absence signals a continuous phase transition.

\begin{figure}[!h]
\centering
\includegraphics[width=0.5\textwidth]{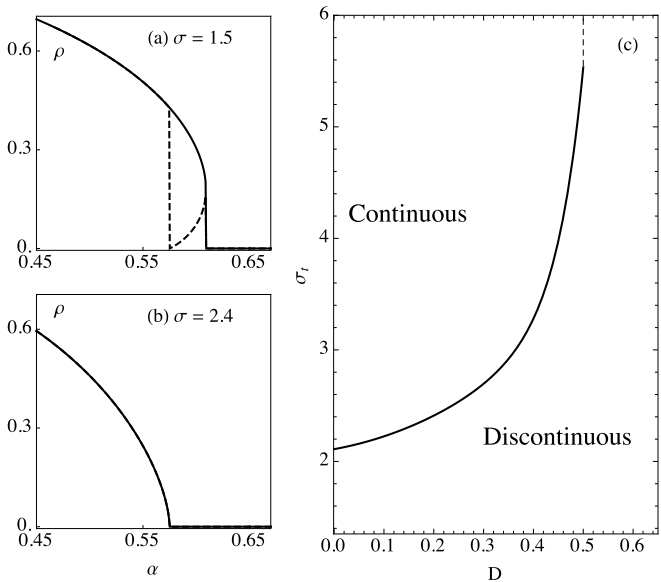}
\caption{ Mean field results. (a) and (b): $\rho$~vs.~$\alpha$ for $a = 2$, $D = 0.15$ and $\sigma = 1.5$ and $2.4$. The dashed curve in (a) correspond to the unstable solution. (c) Phase diagram showing the value $\sigma_t (D)$ where the transition changes from continuous to discontinuous. For $D > 0.5$ the transition is always discontinuous. 
\label{figmf}}
\end{figure}

In Figs.~\ref{figmf} (a) and (b) we exemplify mean-field results
for $a = 2$, $D = 0.15$ and distinct $\sigma$'s. 
The transition is seen to be discontinuous for $\sigma = 1.5$ and continuous for $\sigma = 2.4$.
The complete phase diagram is shown in Fig. \ref{figmf} (c)
along with the tricritical line $\sigma_t(D)$
that separates continuous from discontinuous transitions. 
We see that $\sigma_t$ increases monotonically with $D$ until $D = 0.5$. 
Above this point the transition is always discontinuous. 
This monotonic increase of $\sigma_t(D)$ with $D$, as will be seen below, is in disagreement 
with the numerical simulations. The reasons for this will be discussed in Sec.~\ref{sec:disc}.

%---------------------------------------------------------
\section{Numerical results}

We now present the numerical simulations of the diffusive $\sigma$-CP model. For completeness, and in order to obtain a global picture, we perform simulations in both the constant particle number ensemble and the constant rate ensemble. 

\subsection{Constant particle number  ensemble}

In the constant particle number (conserved) ensemble, 
the total particle number $n$ is held fixed and
 both  creation and  annihilation processes are replaced
by just one jump process. Letting $w_{ij}$  be the transition rate denoting
the jump from a particle at $i$ to an empty site $j$, we have that
\begin{equation}                                                              
w_{ij}= w_i^a w_j^c.                                 
\end{equation}
This jump process can be viewed as the annihilation of a particle 
in  site $i$ and the creation in site $j$.
In Refs.~\cite{hilh02,oliv03,fiore05} it has been shown that
the  mean annihilation rate ${\bar \alpha}$ is given by
\begin{equation}
{\bar \alpha}=\frac{\langle w_j^c\rangle_c }{\langle w_i^a\rangle_c},
\label{eqc}
\end{equation}
where $\langle...\rangle_c$ denotes the average of a given quantity
in this ensemble. Note that the  above expression for the average 
${\bar \alpha}$ is equivalent to
the expression $\langle \omega_j^c \rangle=\alpha \langle w_j^a\rangle$  
obtained
for the constant rate ensemble in the steady regime.
%In both cases, the quantity $\langle \omega_j^c\rangle$ increases 
%by raising $D$. 

An advantage of using the conserved ensemble is that, 
in this version the $\sigma-$CP  does not have, strictly speaking, 
an absorbing state (in contrast to the constant rate ensemble),
except the trivial case $n=0$. 
 Another advantage is that both the transition point and the classification of  the
 transition are readily obtained by performing
numerical simulations for  distinct $n$'s in the subcritical
regime. According to 
Broker and Grassberger \cite{brok99} and 
afterwards \cite{tome01,fiore05,fiore07}, the addition of particles 
placed in an infinite lattice drives  the
system toward the transition point $\alpha_0$ according to the expression
 \cite{brok99,tome01,fiore07},
\begin{equation}                                                              
{\bar \alpha}-\alpha_0 \sim n^{-1}.                                           
\label{eq55}                                                                  
\end{equation}
%Hence we may obtain the transition point $\alpha_0$ by plotting $\bar{\alpha}$~vs.~$1/n$ in the subcritical regime. 
Thus,  we may locate the transition point by linearly extrapolating $1/n$ 
(note that $n$  is held fixed but the system density $\rho \rightarrow 0$).

The classification of
the  phase transition is obtained by  measuring
 the particle  displacements for different $n$.
 Letting $R$ be  the mean distance  between the particles located at the extremities
of the system, we have that \cite{marr99,fiore05,fiore07}
\begin{equation}\label{R}                                                             
R \sim n^{1/d_F},                                                              
\label{fractal}                                                               
\end{equation}
where $d_F$ is the fractal dimension.
 For one-dimensional
systems belonging to the DP universality
class, the clusters are fractals with fractal dimension
 $d_F = 0.74792...$ \cite{vics92}, whereas  at the phase coexistence
it is the proper euclidean dimension $d=1$,
consistent with the emergence  of  a compact cluster.
Hence, the order of the transition may be inferred 
by analyzing the slope of $\ln R$~vs.~$\ln n$.

The actual numerical simulation in the conserved
ensemble is realized as follows. With probability
$D$ a randomly chosen particle hops to its nearest neighbor site (provided
it is empty), whereas with probability $1-D$ the jumping
process is chosen instead. In this case an occupied site is 
chosen at random. If its neighbor is empty, we occupy it 
with  probability
$p_\ell=(1+a\ell^{-\sigma})/(1+a)$, where $\ell$ is the
length of the island of inactive sites in which the active site
is located. The constant factor $1/(1+a)$ is used in order to guarantee
 that $p_\ell\leq 1$.  The particle that occupies the empty site 
is also chosen at random, thus conserving the total particle number $n$.

\begin{figure}
\centering
\includegraphics[width=0.5\textwidth]{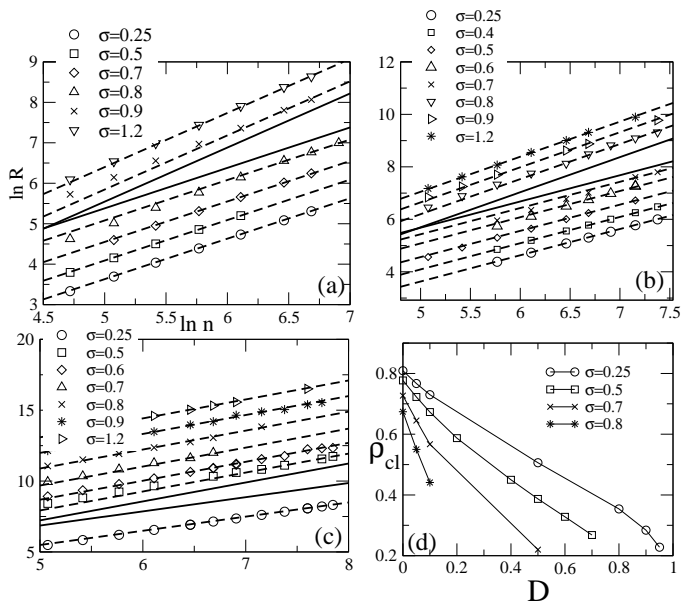}
\caption{Log-log plot of the maximum distance $R$ vs the particle              
number $n$ for distinct values of $\sigma$ and for                             
 $D=0.1$ $(a)$,  $0.5$ $(b)$ and  $0.95$ $(c)$. The upper and
lower straight lines have slopes $1/d_F=$ $1.337...$ and $1$, respectively (cf. Eq.~(\ref{R})).
In (d) we plot the cluster density $\rho_{cl}$ vs $D$, at the phase coexistence,
for distinct values of $\sigma$.}
\label{fig1}
\end{figure}

In Figs. \ref{fig1}  we show results for
$D=0.1$, $0.5$ and $0.95$ and distinct values of $\sigma$.
For $D=0.1$ (Fig. \ref{fig1}(a))  the phase transition is
 discontinuous for $0<\sigma<0.9$ and continuous
for $\sigma \ge 0.9$. This is close to the case $D=0$,
in which the crossover occurs at $\sigma_t=1.0(1)$ \cite{gine05,fiore07}.
The cases  $D = 0.5$ and $0.95$  
exhibit a similar behavior. However,  the effect of diffusion
is now more pronounced with the crossover occurring  
at $\sigma_t = 0.75 (5)$ and
$0.45(5)$ respectively. 
Inspection of the cluster density  $\rho_{cl}=n/R$
(at the phase coexistence) shown in Fig.~\ref{fig1}(d)
reveals that the particle clusters  
become less compact as $D$ increases. 

\begin{figure}
\centering
\includegraphics[width=0.375\textwidth]{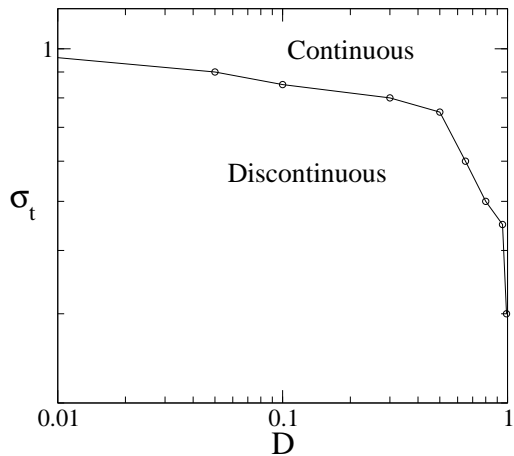}
\caption{The tricritical line $\sigma_t(D)$ obtained from  numerical simulations (cf. Fig.~\ref{fig1}).}
\label{fig12}
\end{figure}

The values of the tricritical line $\sigma_t(D)$ are summarized in Fig.~\ref{fig12}, 
where the monotonically decreasing behavior is clearly observed. 
This is in stark  disagreement with the mean-field predictions of Fig.~\ref{figmf}(c), 
in which $\sigma_t(D)$ grows with $D$. 
The reasons for this discrepancy  will be discussed  below.

 All these
results are found to be similar to those obtained in Ref. \cite{fiore2013},
in which distinct interaction rules have been considered
in order to study the phase coexistence
by ``weakening'' the long-range interaction \footnote{ The
procedures to weaken
the long-range interaction were to 
decrease the long-range parameter $a$ and
the introduction of a competition with
short-range interactions. More details can
be seen in Ref. \cite{fiore2013}.}. Thus, the
above results suggest that the diffusion also  ``weakens'' 
the long-range interaction, in similarity
to Ref. \cite{fiore2013}. As a consequence, 
in a high diffusion regime, 
the phase coexistence yields only for sufficiently  lower $\sigma$'s.
Since the tricritical line $\sigma_t(D)$ is a decreasing function of $D$, 
when $\sigma>1$ the role of the diffusion is irrelevant 
with respect to the change  in  the order of the
transition. In other words, the diffusion
does not shift the phase transition, remaining 
continuous for all values of $\sigma >1$. 
This can be understood by recalling that, when $\sigma$ is large,
the  long-range  factor  $1+a\ell^{-\sigma}$, 
responsible for the occurrence of a phase coexistence, 
decays rapidly with $\ell$, 
 becoming closer to the 
 short-range value $1$. Since in this case the
 phase transition is continuous  for all diffusion rates 
\cite{fiore04}, the conclusion follows.

\subsection{Constant rate ensemble}

In order to confirm the  above conclusions, we have also performed 
numerical simulations
in the constant rate ensemble. 
Unlike the conserved ensemble, the creation and annihilation
rates are the control parameters, whereas the particle density is
a fluctuating quantity.
To locate the transition point $\alpha_0$ and  classify  the 
transition, we perform spreading simulations starting from an initial
configuration with a single particle at the origin. The proper
quantities to evaluate are the survival probability $P_s(t)$, the mean particle
number $N(t)$ and the mean square displacement $R^2(t)$. At the transition
point they follow power-law behaviors
given by 
\begin{equation}\label{exponents}
P_s(t) \sim t^{-\delta},\qquad  N(t) \sim t^{\eta}, \qquad R^{2}(t) \sim t^{2/z},
\end{equation}
where $\delta, \eta$ and $z$ are associated
dynamic critical exponents. For continuous transitions belonging 
to the DP universality class these exponents present the well
known values   
\begin{equation}\label{expo1}
\delta=0.159464(6), \quad \eta=0.313686(8), \quad z=1.580 745(10).
\end{equation}
Instead, at the one-dimensional phase coexistence (despite the order-parameter
gap) their values read \cite{henkel,odor04}
\begin{equation}\label{expo2}
\delta = 1/2, \quad \eta = 0, \quad z = 1.
\end{equation}
Hence, the order of the phase transition 
can be obtained from the values of
these critical exponents. 

This analysis is also useful since it 
allows to draw  a comparison with results
obtained from the conserved ensemble, whose 
above dynamic exponents and  fractal dimension $d_F$ are related through 
the expression $d_{F}=2(\eta+\delta)/z$. Away from the critical
point, all quantities deviate from power-law behaviors,
reaching a regime of endless activity for $\alpha<\alpha_0$ and exponential decay toward
extinction for   $\alpha>\alpha_0$. 
The continuous transitions have also been confirmed by studying
the time decay of the system density $\rho$ starting from a fully
occupied initial configuration. At the critical point it behaves
as $\rho(t) \sim t^{-\theta}$, where for the CP $\theta=\delta=0.159464(6)$. 
In contrast, in the active and absorbing
phases, $\rho(t)$  converges to
a  well defined value ${\bar \rho} \neq 0$ 
and  vanishes exponentially, respectively.
For the evaluation
of $\rho$, we have considered $L=20000$ and averages have
been evaluated over
$30000$ initial configurations.
%For distinct values of $\sigma$ I will take two values
%of diffusion, each one exemplifying the regime of
%low and large diffusion rates.  The transition point is obtained  
%by inspecting an algebraic
%behavior of $N$ vs $P_s$,
%where they behave according to relation $N \sim P_{s}^{-\eta/\delta}$. 
%At a one-dimensional DP critical
%point the ratio $-\eta/\delta$ read $1.9658...$, 
%whereas at the phase coexistence it reads $0$.
%Starting with $\sigma=0.8$ (where the $D=0$ case presents 
%a discontinuous transition), 

\begin{figure}
\centering
\includegraphics[width=0.5\textwidth]{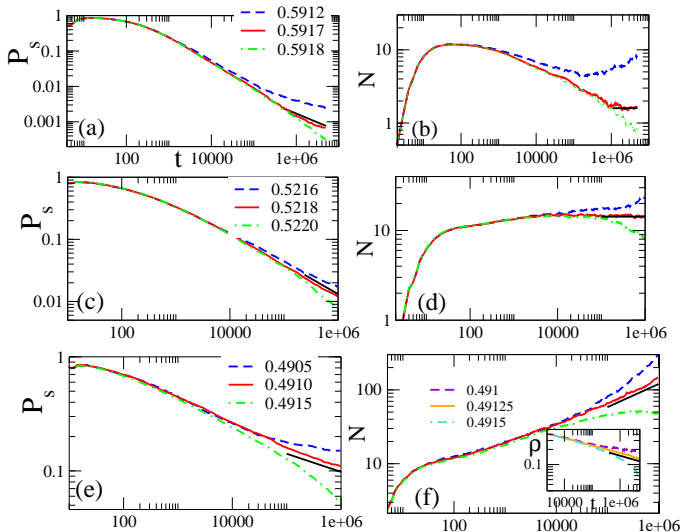}
\caption{({\bf color online}) The time evolution of $P_s(t)$ and $N(t)$ in the constant rate ensemble for distinct values of $\alpha$ and  $D=0.1$. (a)-(b) $\sigma=0.25$; (c)-(d) $\sigma=0.8$; (e)-(f) $\sigma = 1.2$. The slopes of the black straight lines are given by Eq.~(\ref{expo2}) for (a)-(d) and Eq.~(\ref{expo1}) for (e)-(f).
 The inset in image (f) shows the decay of the density $\rho$ 
and the black line has slope $\theta=0.159464(6)$. }
\label{fig2}
\end{figure}

\begin{figure}
\centering
\includegraphics[width=0.5\textwidth]{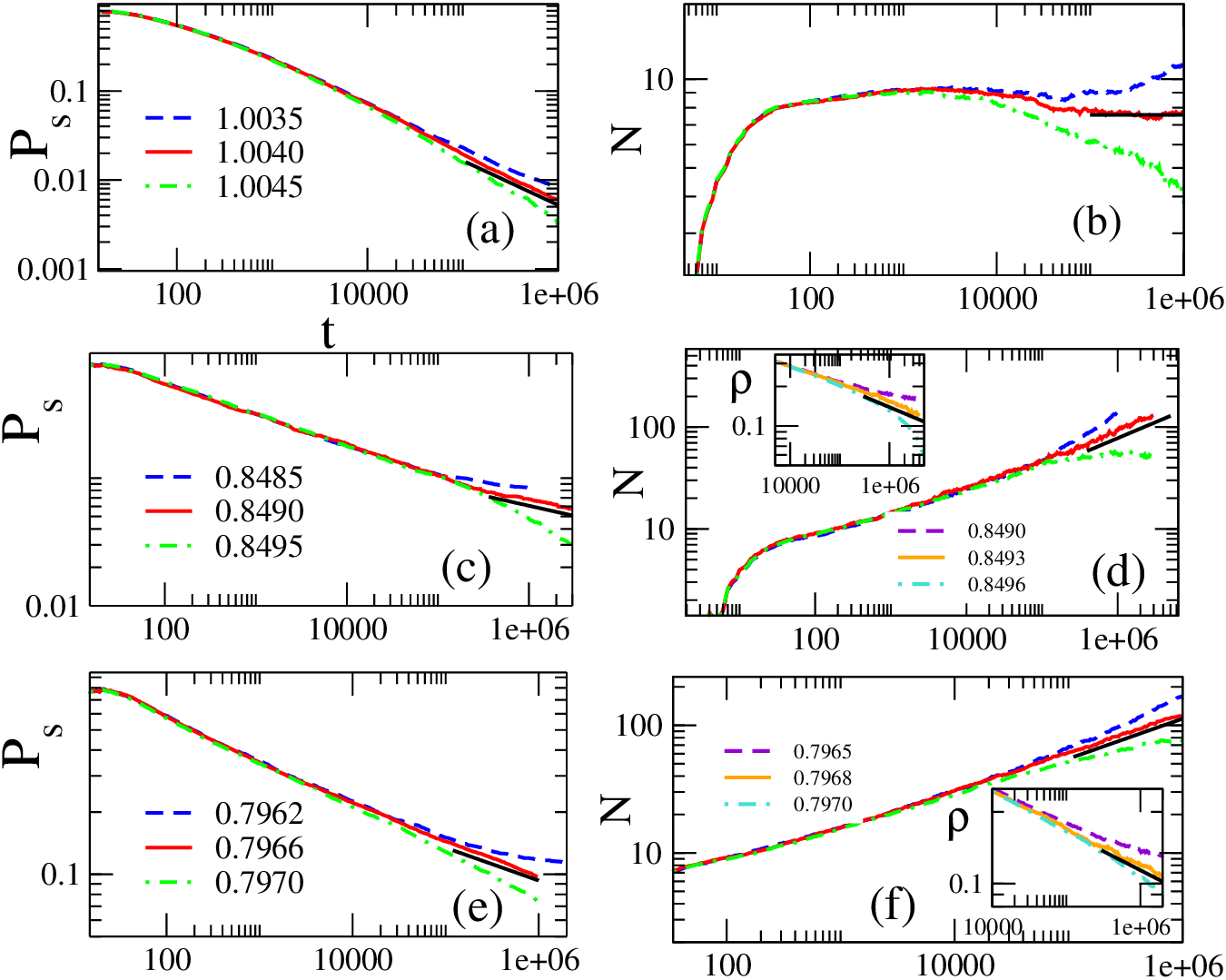}
\caption{({\bf color online}) Same as Fig.~\ref{fig2} but for $D = 0.5$. 
The slopes of the black straight lines are given by Eq.~(\ref{expo2}) for (a)-(b) and Eq.~(\ref{expo1}) for (c)-(f). The insets show the time decay of
$\rho$ for distinct $\alpha$'s and the black lines have slope 
$\theta=0.159464(6)$.}
\label{fig3}
\end{figure}

\begin{figure}
\centering
\includegraphics[width=0.5\textwidth]{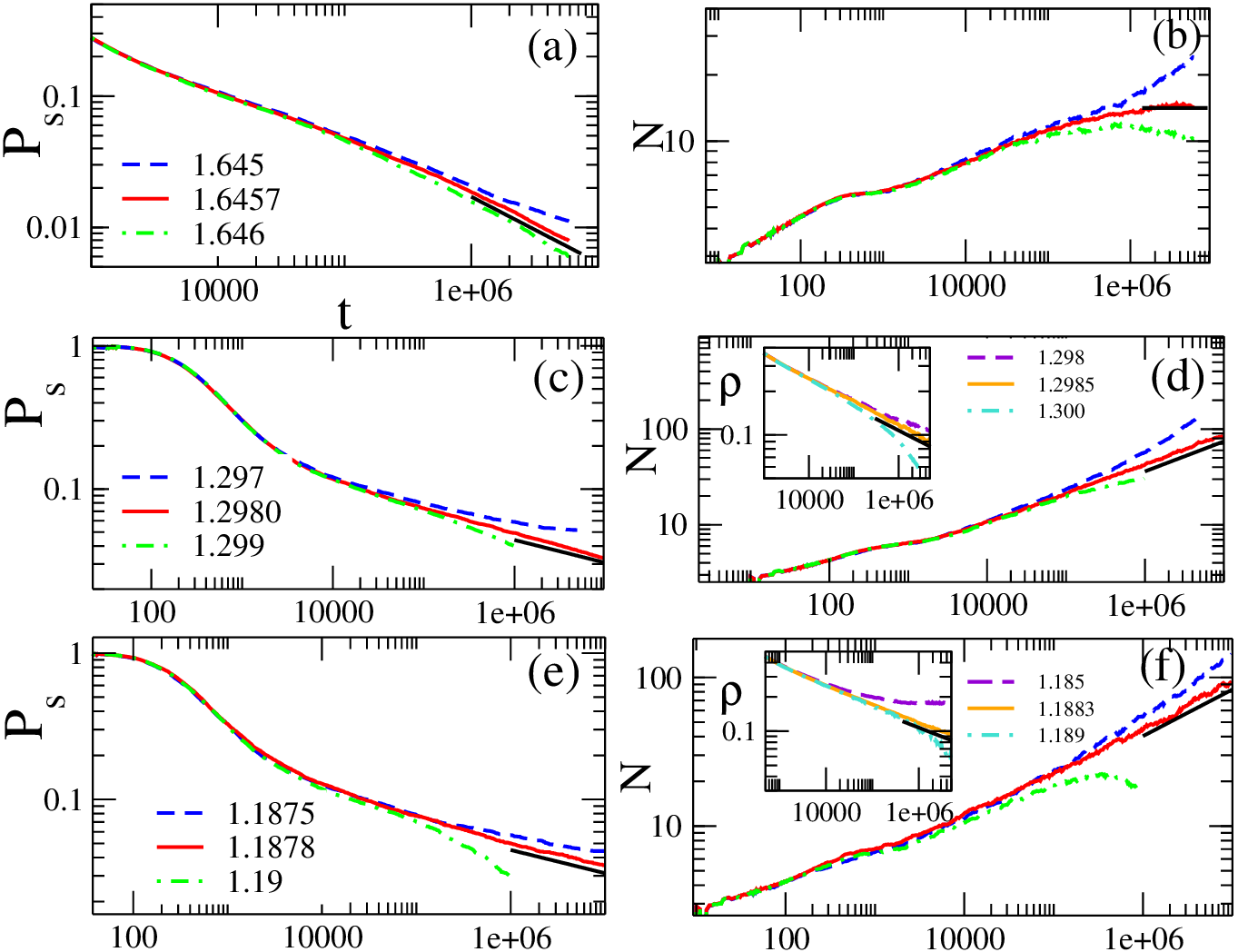}
\caption{ ({\bf color online}) Same as Fig.~\ref{fig3} but for $D = 0.95$. The slopes of the black straight lines are given by Eq.~(\ref{expo2}) for (a)-(b) and Eq.~(\ref{expo1}) for (c)-(f). The insets show the time decay of         
$\rho$ for distinct $\alpha$'s and the black lines have slope                
$\theta=0.159464(6)$.}
\label{fig4}
\end{figure}

The main results for the constant rate ensemble are shown in Figs.~\ref{fig2}-\ref{fig4} for $D = 0.1$, $0.5$ and $0.95$ respectively. In each figure we plot both $P_s(t)$ and $N(t)$ for three distinct values of $\sigma$; namely $0.25$, $0.8$ and $1.2$. In all cases, above quantities have been calculated
over $30000$ initial configurations. 
In Fig. \ref{fig2} we show the main results for  $D=0.1$. 
When $\sigma=0.25$ and $0.8$ (figures (a)-(d)), we observe 
an algebraic behavior  with the exponents
 $\delta=1/2$ and $\eta=0$  of Eq.~(\ref{expo2}) (see also Eq.~(\ref{exponents})). 
 This signifies, in agreement with Fig.~\ref{fig1}, a discontinuous transition. 
On the other hand,  for $\sigma=1.2$ (figures (e) and (f)) the 
power-law regime has the DP exponents of Eq.~(\ref{expo1}), indicating a continuous transition. The time decay of $\rho$,
shown in the inset of Fig.~\ref{fig2} (d), 
confirms the algebraic behavior with the DP exponent.
%at $\alpha_0=0.49125(25)$ (this estimate is somewhat larger than that obtained from  spreading experiments).  
The analysis is repeated in Figs.~\ref{fig3} and \ref{fig4} for $D = 0.5$ and $0.95$. In both cases  the phase transition is discontinuous 
for $\sigma = 0.25$ and  continuous for $\sigma = 0.8$ and $1.2$, in agreement
with results from the conserved ensemble and thus
confirming that $\sigma_t$ decreases by raising $D$. Analysis
of the time decay of $\rho$  reinforce this results. The middle curves in
the insets  present algebraic decays 
 consistent with the DP values $\theta=0.159464(6)$
at $\alpha$'s very close to the  $\alpha_0$. In fact,  
these estimates  are somewhat larger than $\alpha_0$, due to finite size effects.

Hence, we conclude that the results of the constant rate ensemble are
in complete agreement with those of the conserved ensemble. 
In contrast with mean-field results,  the value $\sigma_t$, where
the order of the transition changes, diminishes with increasing diffusion.

\section{\label{sec:disc}Discussion and Conclusion}

In this paper we have investigated the role of diffusion in 
the simplest model presenting a discontinuous phase transition with an absorbing state. It is a counterpart of the usual contact process, in which 
the particle creation rate depends on the length $\ell$ of inactive islands surrounding the 
creation site according to $1 + a \ell^{-\sigma}$. 
In the absence of diffusion, a tricritical point $\sigma_t$
separating discontinuous from continuous transition. For $a = 2$ this point occurs at $\sigma_t = 1.0(1)$.

We investigated, by means of numerical simulation and mean-field calculations,  
the effect of diffusion on the phase coexistence regimes and over $\sigma_t$.
Results for distinct values of $\sigma$ and diffusion rates showed that 
 the crossover $\sigma_t$ is monotonically reduced as the diffusion increases, similarly to the
 interactions introduced to weaken the long-range feature studied in Ref.~\cite{fiore2013}. 
%In particular, $a = 2$, we found $\sigma_t (0) = 1.0(1)$,  $\sigma_t (0.95) = 0.5(1)$.
This suggests that by 
increasing the diffusion toward the limit $D\to 1$, only sufficient small $\sigma$ 
are able to stabilize compact clusters. In fact, results 
for $D=0.99$ (cf Fig.~\ref{fig12}) show that for sufficiently 
low $\sigma$ ($\sigma \le 0.2$), the transition is still discontinuous, but
the value $\sigma_t \sim 0.3$ signals the emergence of a continuous transition. 
Hence, our results reveal a novel role played by the diffusion in phase transitions, 
being responsible for weakening
the compact displacement among the particles. 
Notwithstanding, we  emphasize that  further studies of discontinuous absorbing transitions in the presence of diffusion are necessary in order to yield a complete picture of the problem.

Finally, we turn to the marked disagreement between the mean-field results and the numerical simulations, regarding the dependence of $\sigma_t$ on $D$. As we have seen, in the mean-field approximation $\sigma_t$ was found to be a monotonically increasing function of $D$, whereas in the numerical simulations the exact opposite behavior was observed. Moreover, in the mean-field we have found that above $D=0.5$ the transition should  be discontinuous for any value of $\sigma$. 
We attribute this disagreement to the correlations neglected by
the mean-field theory, that become more important in low dimensions 
(as in the present case).
 In fact a similar disagreement between
mean-field and  numerical simulations
in an one-dimensional discontinuous transition has been
recently investigated in Ref. \cite{dic09}.
 Another possibility concerns the  restrictions of the lattice
particle occupations   (only one particle can occupy a given site)
that, together with the present lattice topology, could
 prevent the  particle clustering induced by increasing the diffusion. In other words,
a lattice model allowing multiple occupation of each site  
may lead to results compatible with
mean-field predictions. 
We remark that, notwithstanding,  all these points deserve further investigations. 

\section*{ACKNOWLEDGEMENT}
C. E. F. acknowledges the financial support from   CNPQ 
and G. T. L. acknowledges the support from FAPESP.


\begin{thebibliography}{99}




\bibitem{marr99}
J. Marro and R.Dickman,
{\it Nonequilibrium Phase Transitions in Lattice Models}
(Cambridge University Press, Cambridge, England, 1999).


\bibitem{ziff86}
R. M. Ziff, E. Gulari and Y. Barshad,
Phys. Rev. Lett. {\bf 56}, 2553 (1986).

\bibitem{fiore04}
C. E. Fiore and M. J. de Oliveira,
Phys. Rev. E {\bf 70}, 046131 (2004).



\bibitem{hinr00}
H. Hinrichsen, cond-mat/0006212.

\bibitem{gine05}
F.Ginelli, H. Hinrichsen, R. Livi, D. Mukamel and A. Politi,
Phys. Rev. E {\bf 71}, 026121 (2005).

\bibitem{harr74}
T. E. Harris, Ann. Probab. {\bf 2}, 969 (1974).

\bibitem{fiore2013} C. E. Fiore and M. J. de Oliveira, Phys. Rev. E 
{\bf 87}, 042101 (2013).

\bibitem{fiore2014} C. E. Fiore, Phys. Rev. E {\bf 89}, 022104 (2014).

\bibitem{vojta} See for example, T. Vojta and M. Dickison, Phys. Rev. E
{\bf 72},  036126 (2005); H. Barghathi and T. Vojta, 
Phys. Rev. Lett {\bf 109}, 170603 (2012).

\bibitem{henkel} M. Henkel, H. Hinrichsen and S. Lubeck, {\it
Non-Equilibrium Transitions, Volume 1} (Springer, 2008).
\bibitem{odor04} G. Odor, Rev. Mod. Phys {\bf 76}, 663 (2004).


\bibitem{munoz2014} P. V. Martin, 
J. A. Bonachela and M. A. Munoz, Phys. Rev. E {\bf 89}, 012145 (2014).

\bibitem{dic89} R. Dickman, Phys. Rev. B {\bf 40}, 7005 (1989).



\bibitem{tome01} T. Tom\'e and M. J. de Oliveira, Phys. Rev.
Lett. {\bf 86}, 5643 (2001).

\bibitem{hilh02}
H. J. Hilhorst and F. van Wijland,
Phys. Rev. E {\bf 65}, 035103 (2002).

\bibitem{oliv03}
M. J. de Oliveira,
Phys. Rev. E {\bf 67}, 027104 (2003).

\bibitem{fiore05}
C. E. Fiore and M. J. de Oliveira,
Phys. Rev. E {\bf 72}, 046137 (2005).

\bibitem{ziff92}
R. M. Ziff and B. J. Brosilow, 
Phys. Rev. A {\bf 46}, 4630 (1992).


\bibitem{fiore07} C. E. Fiore and M. J. de Oliveira,
Phys. Rev. E {\bf 76}, 041103 (2007).


\bibitem{losc02}
E. S. Loscar and E. V. Albano 
Phys. Rev. E {\bf 65}, 066101 (2002).




\bibitem{brok99}
H.-M. Br\"oker and P. Grassberger,
Physica A {\bf 267}, 453 (1999).

\bibitem{vics92}
T. Vicsek, {\it Fractal Growth Phenomena}, 2nd ed.
(World Scientific, Singapoure, 1992).

\bibitem{dic09} G. Odor and R. Dickman, J. Stat. Mech. {\bf 2009}, p08024
(2009).

\end{thebibliography}
\end{document}